*RESEARCH ARTICLE*

# A PATH ANALYSIS OF COVID-19 WITH THE INFLUENCE OF AIR PRESSURE, AIR TEMPERATURE, AND RELATIVE HUMIDITY

**Marvin G. Pizon, Ronald R. Baldo, Ruthlyn N. Villarante and Jessica D. Balatero**
Agusandel Sur State College of Agriculture and Technology.

……………………………………………………………………………………………………....

## *Manuscript Info*

## *Abstract*

Coronavirus disease 2019 (COVID-19) is one of the most infectious diseases and one of the greatest challenge due to global health crisis. The virus has been transmitted globally and spreading so fast with high incidence. While, the virus still pandemic, the government scramble to seek antiviral treatment and vaccines to combat the diseases. This study was conducted to investigate the influence of air pressure, air temperature, and relative humidity on the number of confirmed cases in COVID-19. Based on the result, the calculation of reproduced correlation through path decompositions and subsequent comparison to the empirical correlation indicated that the path model fits the empirical data. The identified factor significantly influenced the number of confirmed cases of COVID-19. Therefore, the number of daily confirmed cases of COVID-19 may reduce as the amount of relative humidity increases; relative humidity will increase as the amount of air temperature decreases; and the amount of air temperature will decrease as the amount of air pressure decreases. Thus, it is recommended that policy-making bodies consider the result of this study when implementing programs for COVID-19 and increase public awareness on the effects of weather condition, as it is one of the factors to control the number of COVID-19 cases.

……………………………………………………………………………………………………....

## Introduction:-

A sudden outbreak of coronavirus disease 2019 (COVID-19) happened since December 2019 in the city of Wuhan, China. The virus has spread to several other countries around the world and has been identified as the cause of an outbreak of respiratory illness, with symptoms such as fever, cough, shortness of breath, muscle ache, confusion, headache, sore throat, rhinorrhoea, chest pain, diarrhea, nausea, and vomiting (Hui et al., 2020; Chen et al., 2020). As of April 11, 2020, a total of 1,524,161 confirmed cases reported in 213 countries and it was one of the biggest infectious disease outbreak worldwide, and biggest battle since the disease is spreading so fast with high incidence and the transmission has evolved all people in the country (WHO, 2020).

In Philippines contexts, around first week of the March 2020, the Department of Health (DOH) reported one confirmed cases in COVID-19. As of April 11, 2020, it has increased to 4428 cases in all provinces. Although, government has taken various measures to control the spread of COVID-19, like implementing social distancing, self-quarantine and advising people to practice proper hygiene to prevent them from being infected. Also,

**Corresponding Author:-Marvin G. Pizon**
Address:- Agusandel Sur State College of Agriculture and Technology.

researchers doing their best to help the government reduce the number of daily confirmed COVID-19 cases through uncontrolled factors.

The transmission ofcoronaviruses can be affected by a several factors, such as; temperature and humidity, population density and medical care quality, including climate conditions (Hemmes, 1960; Dazkiel, 2018).Therefore, understanding the relationship between weather conditions is an important tool to forecast the number of confirmed COVID-19 cases.Previous studies have shown the importance of weather variables in the transmission of infectious diseases. It has been recorded that the number of the newly confirmed cases of COVID-19 increased in places having low temperatures and low humidity, while much fewer cases were recorded in places with high temperature and high humidity (Bukhari and Jameel, 2020). This result was consistent to the study conducted by Tang et. al (2020), that high temperature and high humidity significantly reduce the transmission of COVID-19.

Several mathematical models have been proposed to describe the transmission of COVID-19 through the use of weather condition. However, due to limited emerging understanding of the new virus and its transmission mechanisms, the results are focusing only two parameters of weather conditions. Thus, this study has been formulated to forecast the number of confirmed COVID-19 cases through three weather parameters. The main objectives of this study is to identify effects of causal ordering of weather condition such as; air pressure, air temperature, and relative humidity on infected areas. The study generated the model through Path Analysis in order to establish the specific cause-and-effect among air pressure, air temperature, and relative humidity. Further, the created model was tested over diagnostic checking on the underlying assumptions for its robustness.

**Literature Review And Building Of Path Diagram:**

The exogenous variables of this research were Air Pressure ($X_1$), Air Temperature ($X_2$), and Relative Humidity ($X_3$). Meanwhile, the number of confirmed COVID-19cases (Y) was identified as the endogenous variable. This study had looked into the potentials of using these variables based on the following literature reviews:

Based on the study of Tosepu and Asfian (2020), weather temperature is one of the factors that triggered the spread of COVID-19. This result was in accordance to the study of Ud-Dean(2010) that environmental factors including relative humidity, vapor pressure and temperature are known to affect seasonal virus survival and transmission. These results were consistent with the study of Wang et al. (2020) and Ma et. al. (2020), that high temperature and high humidity reduce the transmission of COVID-19. Also, the study of Chen et. al. (2020) stated that the population that emigrated where the virus was epidemic on that particular place was the main infection source in other cities and provinces.

Several studies point out that relative humidity was dependent on air temperature. The study of Pfahl and Niedermann (2011) found out that there was a negative strong correlation between temperature and relative humidity. Ajadi and Sanusi (2013) also point out that the higher the temperature the lower the relative humidity and hence the faster the drying rate of any material and humid air slows down evaporation.

Air temperature and relative humidity are important property of a climate on a specific locations. The study of Wooten (2011) and Dines (1925), revealed that air pressure was highly correlated to atmospheric temperature. Consequently, air temperature differences between different locations will cause air pressure differences, which in turn would produce air movement. Also, the study revealed that the moisture of holding capacity of air depends on the air's temperature, as the moisture holding capacity increases the relative humidity decreases.

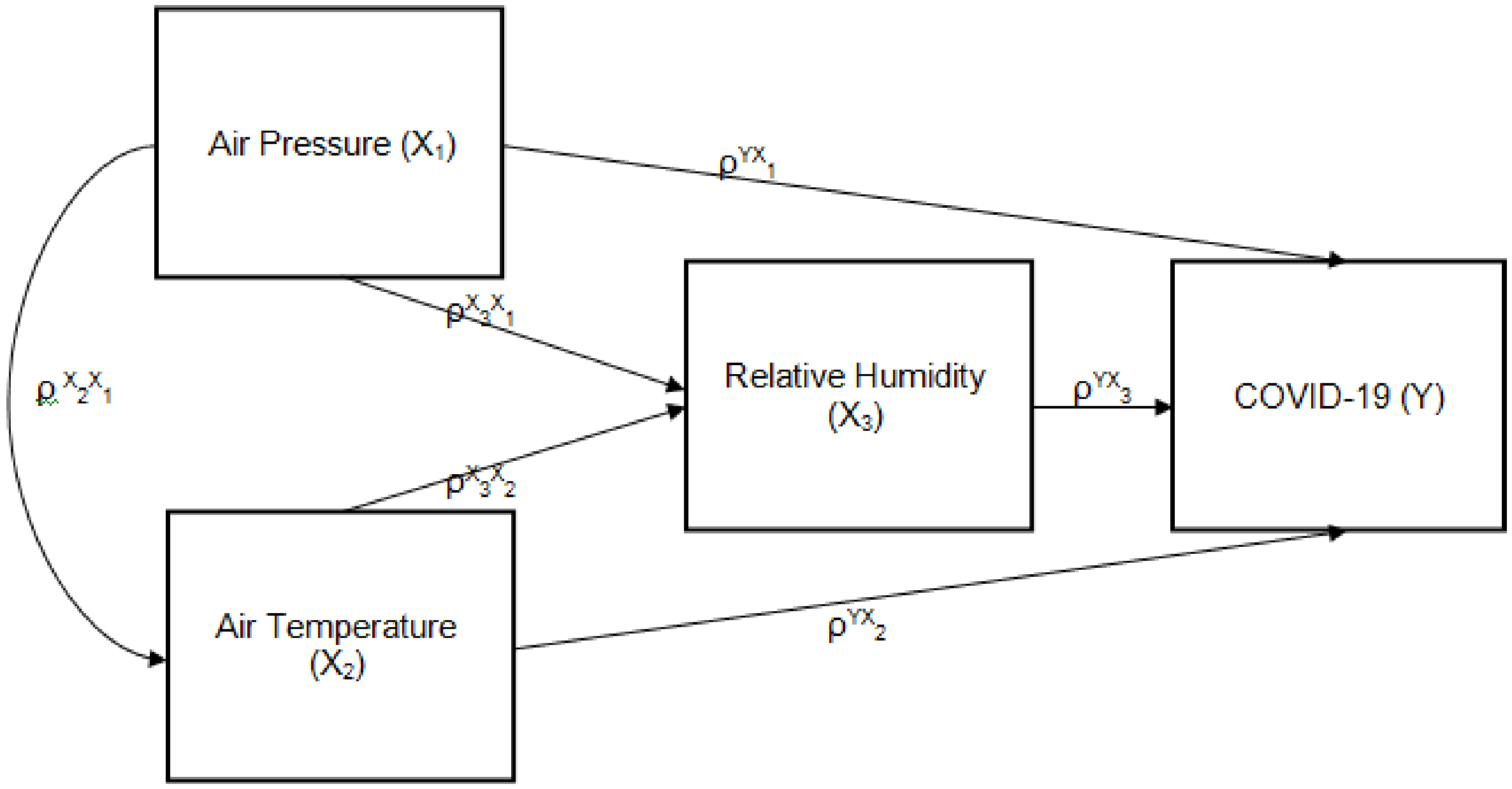

**Figure 1:-** Input Path Diagram Representing a Proposed Causal Model.

Legend: $\rho^{YX_1}$ – Path coefficient influence of Air Pressure towards COVID-19
$\rho^{YX_2}$ – Path coefficient influence of Air Temperature towards COVID-19
$\rho^{YX_3}$– Path coefficient influence of Relative Humidity towards COVID-19
$\rho^{X_2X_1}$ – Path coefficient influence of Air Pressure towards Air Temperature
$\rho^{X_3X_1}$ – Path coefficient influence of Air Pressure towards Relative Humidity
$\rho^{X_3X_2}$ – Path coefficient influence of Air Temperature towards Relative Humidity

The causal model in Figure 1 proposed that the number of confirmed cases of COVID-19 results from theair pressure, air temperature, and relative humidity from different region.Furthermore, the path analysis was carried out through multiple linear regression procedure. The causal model will be predicting the direct and indirect effect.

In this study, the model is specified by the following path equations:
$Y = \rho^{YX_1}X_1 + \rho^{YX_2}X_2 + \rho^{YX_3}X_3 + e_1$ (1)
$X_3 = \rho^{X_3X_2}X_1 + e_2$ (2)
$X_2 = \rho^{YX_1}X_1 + e_2$ (3)
where:
Y = COVID-19 confirmed cases
$X_1$ = Air Pressure
$X_2$ = Air Temperature
$X_3$ = Relative Humidity

## Methodology:-

This study used secondary data obtained from the available online websites: https://www.doh.gov.ph/covid19tracker and https://www.timeanddate.com/weather, which provide the datasets containing the number of confirmed cases of COVID-19 in all affected regions of the Philippines and the data of the three weather parameters (air pressure, air temperature, and relative humidity), respectively.

Around first week of March 2020, the Philippines Department of Health (DOH) confirmed the first local transmission of COVID-19. Hence, the researchers considered the data starting from the first confirmed case up to April 13, 2020. Furthermore, the researchers considered average data of weather conditions starting from March 2020 up to April 13, 2020.

The researchers used the following units on the weather parameters: inch of mercury (inHg and Hg) for air pressure (this was used for barometric pressure in weather reports), degrees Fahrenheit (°F) for air temperature, and percent (%) for the relative humidity (this was used on describing how much amount of water vapor in the air that can hold at the given temperature).

After collecting the data from all sources, the data were checked for further analysis, that is, checked the restriction of range in the data values, outliers, nonlinearity, and non-normality of data, in order to determine the aptness of the generated model.

## Results And Discussion:-

The aptness of the generated model for the said data was evaluated and was tested whether it satisfies the required assumptions. The range of values obtained for variables was considered as a restricted range of one or more variables can reduce the magnitude of relationships. Mahalanobis distance was performed in order to detect the outliers as it can strongly affects the mean and standard deviation of a variable. The linearity assumption was also considered and can best tested with scatter plots, whether the number of confirmed COVID-19cases are linearly related to the air pressure, air temperature and relative humidity.

Moreover, the analysis requires to check that whether all variables were normally distributed since it determined with a goodness of fit test and affects the resulting Path Analysis. The study used the Kolmogorov-Smirnov test in detecting non-normality. Variance Inflation Factor (VIF) values were also apply to test the multicollinearity assumption, this is to check that the independent variables are not highly correlated with each other. And lastly, the assumption of homoscedasticity were considered through the plot of standardized residuals against predicted values, since it allows to test the variance of error terms in the values of the independent variables.After the basic assumptions had been met, data were analyzed through path analysis.

Figure 2 displays results of the initial path analysis model of number of confirmed cases of COVID-19, relative humidity, air temperature, and air pressure. The obtained regression coefficients in Figure 2 was specified by Equation (1– 3) through multiple regression analysis. Among the three exogenous variables only the relative humidity were negatively significant in the number of confirmed COVID-19 cases (-0.521*). Moreover, the air temperature was negatively correlated to relative humidity (-0.670), and air pressure was positively correlated to air temperature (0.804*).

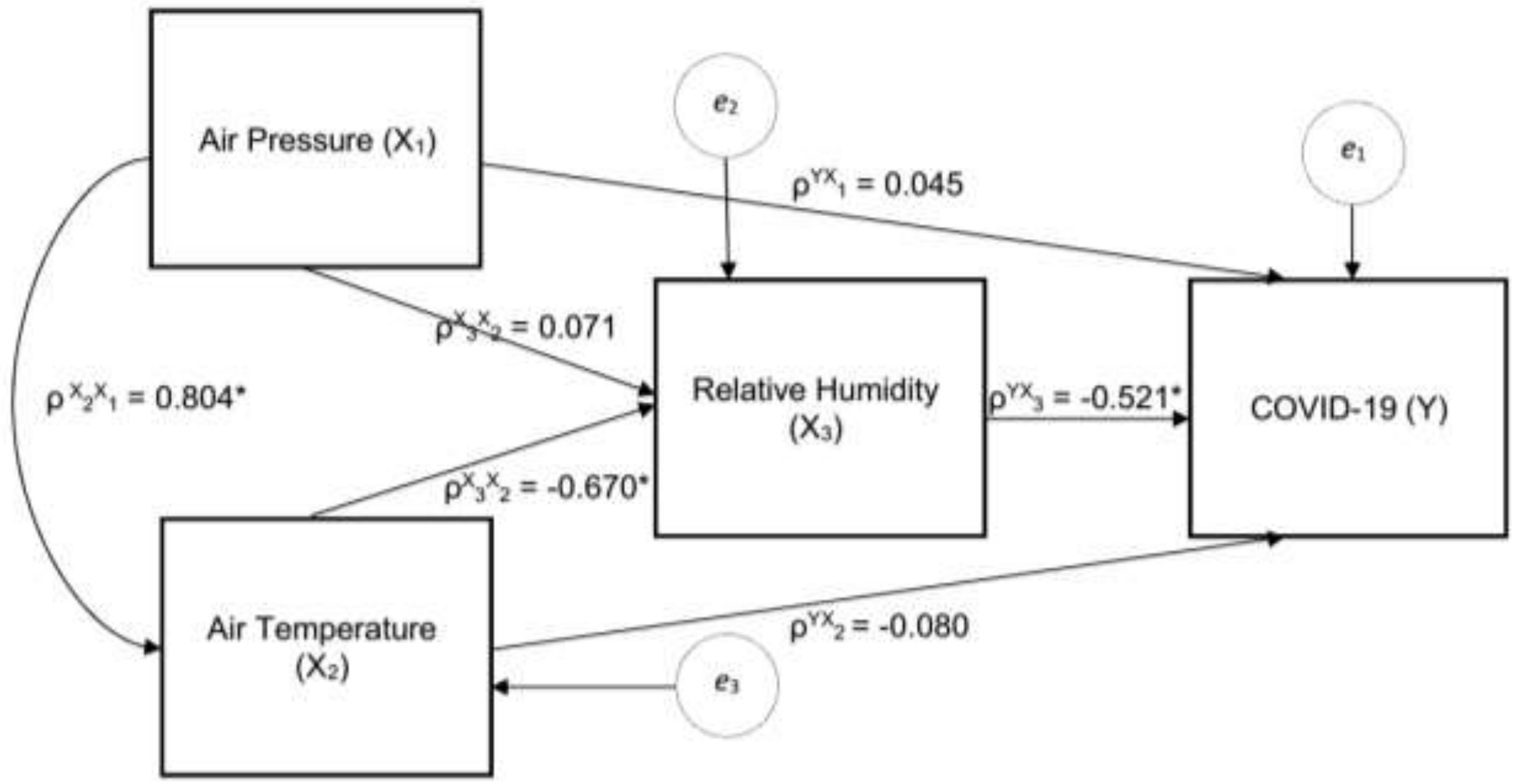

**Figure 2:-** Initial Causal Factor Models Affecting the COVID-19.

Table 1 shows the calculation of observed correlation for the number of confirmed COVID-19 cases model. The magnitude of the Pearson correlation coefficient determines the strength of the correlation. Although there are no concrete rules for assigning the strength of association to particular values, the study had used the general guideline provided by Cohen (1988):

| Coefficient Value | Strength of Association |
|---|---|
| $0.1 < \lvert r \rvert < 0.3$ | Small Correlation |
| $0.3 < \lvert r \rvert < 0.5$ | M<br>edium/Moderate Correlation |
| $\lvert r \rvert > 0.5$ | Large/Strong Correlation |

**Table 1:-** Calculation of Observed Correlation for the COVID-19 Model.

| | | $X_1$ | $X_2$ | $X_3$ | Y |
|---|---|---|---|---|---|
| $X_1$ | Pearson Correlation | 1 | **.804**$^{**}$ | **-.469**$^{**}$ | .225 |
| | Sig. (2-tailed) | | .000 | .001 | .143 |
| | N | 44 | 44 | 44 | 44 |
| $X_2$ | Pearson Correlation | **.804**$^{**}$ | 1 | **-.613**$^{**}$ | .276 |
| | Sig. (2-tailed) | .000 | | .000 | .070 |
| | N | 44 | 44 | 44 | 44 |
| $X_3$ | Pearson Correlation | **-.469**$^{**}$ | **-.613**$^{**}$ | 1 | **-.493**$^{**}$ |
| | Sig. (2-tailed) | .001 | .000 | | .001 |
| | N | 44 | 44 | 44 | 44 |
| Y | Pearson Correlation | .225 | .276 | **-.493**$^{**}$ | 1 |
| | Sig. (2-tailed) | .143 | .070 | .001 | |
| | N | 44 | 44 | 44 | 44 |
| **. Correlation is significant at the 0.01 level (2-tailed). | | | | | |

Based on the table, the relative humidity ($X_3$) is inversely related to COVID-19 (Y), which obtained p-values less than 0.05 level of significance. The strength of association from these variables revealed to have a medium/moderate correlation. This implies that as the amount of relative humidity increases, then the number of confirmed COVID-19 cases decreases.

Meanwhile, air pressure ($X_1$) and air temperature ($X_2$) were inversely related to relative humidity ($X_3$), which obtained p-values less than 0.05 level of significance. The strength of association from $X_1$ to $X_3$ revealed to have a medium/moderate correlation, while the strength of association from $X_2$ to $X_3$ revealed to have a large/strong correlation. These results imply that as the air pressure and air temperature decrease, then the relative humidity will increase.

Lastly, air pressure ($X_1$) was directly related toair temperature($X_2$) which obtained p-value of 0.000. The strength of association from $X_1$to$X_2$reveals large/strong correlation. This indicate that when air pressure decreases, the air temperature will also decrease.

To evaluate the model fit in Figure 2, obtaining the reproduced correlations and comparing it to the empirical correlations must be needed to assess the consistency of the model.To determine the reproduced correlation between two variables, it involves the identification of all valid paths between the variables in the model. The complete set of path decompositions and reproduced correlations for the model shown in Figure 2 is presented in Table 2. Causal effects are presented by paths consisting only of causal links, that is, only one-headed arrow is used to indicate the effect of a variable pressumed to be the cause on another variable pressumed to be an effect. In this study, a direct effect or a causal path consisting of only one link is denoted by “D”, a direct effect consisting of two or more link is denoted by “I”, and spurious effect that is any path components resulting from paths that have reversed casual direction at some point is denoted by “S”.

**Table 2:-** Calculation ofInitial Reproduced Correlation for the COVID-19 Model.

| Reproduced | Path Decomposition |
|---|---|

| Correlation | |
|---|---|
| $\hat{r}_{12}$ | $= \rho^{X_2X_1}$ |
| | = **0.804** |
| | (D) |
| $\hat{r}_{13}$ | $= \rho^{X_3X_1} + (\rho^{X_2X_1})(\rho^{X_3X_2})$ |
| | = 0.71+ ( 0.804)(-0.670) = **0.17132** |
| | (D) (I) |
| $\hat{r}_{1y}$ | $= \rho^{YX_1} + (\rho^{X_2X_1})(\rho^{YX_2}) + (\rho^{X_2X_1})(\rho^{X_3X_2})(\rho^{YX_3}) + (\rho^{X_3X_1})(\rho^{YX_3})$ |
| | = 0.045+ (0.804)(-0.080) + (0.804)(-0.670)(-0.521) + (0.71)(-0.521) = **-0.109** |
| | (D) (I) (I) (I) |
| $\hat{r}_{23}$ | $= \rho^{X_3X_2} + (\rho^{X_2X_1})(\rho^{X_3X_1})$ |
| | = (-0.670)+(0.804)(0.71)= **-0.099** |
| | (D) (S) |
| $\hat{r}_{2y}$ | $= \rho^{YX_2} + (\rho^{X_3X_2})(\rho^{YX_3}) + (\rho^{X_2X_1})(\rho^{X_3X_1})(\rho^{YX_3}) + (\rho^{X_2X_1})(\rho^{YX_1})$ |
| | = (-0.080)+(-0.670)(-0.521)+( 0.804)(0.71)(-0.521)+(0.804)(0.045)=**0.008** |
| | (D) (I) (S) (S) |
| $\hat{r}_{3y}$ | $= \rho^{YX_3} + (\rho^{X_3X_1})(\rho^{YX_1}) + (\rho^{X_3X_1})(\rho^{X_2X_1})(\rho^{YX_2}) + (\rho^{X_3X_2})(\rho^{YX_2})$ |
| | = -0.521+(0.71)(0.045)+(0.71)(0.804)(-0.080)+(-0.670)(-0.080)= **-0.481** |
| | (D) (S) (S) (S) |

To obtain the reproduced correlation (initial model) in Table 3, the set of legitimate paths in Table 2 was used, that is, making the substitutions of path coefficients in Figure 2. In assessing the fit of the model in Figure 2, it can be gleaned from Table 3 that three out of six reproduced correlations have differences greater than 0.05. Hence, those reproduced correlations that have differences greater than 0.05 from the empirical correlations indicate that the model is not consistent with the empirical data.

**Table 3:-** Observed and Initial Reproduced Correlation for the COVID-19 Model.

| | $X_1$ | $X_2$ | $X_3$ | Y |
|---|---|---|---|---|
| Observed Correlation | | | | |
| $X_1$ | 1 | | | |
| $X_2$ | 0.804 | 1 | | |
| $X_3$ | -0.469 | -0.613 | 1 | |
| Y | 0.225 | 0.276 | -0.493 | 1 |
| Reproduced Correlation | | | | |
| $X_1$ | 1 | | | |
| $X_2$ | 0.804 | 1 | | |
| $X_3$ | 0.171 | -0.099* | 1 | |
| Y | -0.109* | 0.008* | -0.481 | 1 |

*Difference between reproduced and observed correlation is greater than 0.05.

Upon examining the significance tests for several paths, the beta coefficient from $X_1$ to Y ($\rho^{YX_1}$), $X_2$ to Y ($\rho^{YX_2}$), and $X_1$ to $X_3$ ($\rho^{X_3X_1}$)was not significant. It suggests that these paths should be removed and revisions to the model are warranted prior to describing any of the causal effects. The revised path diagram, including path coefficients is presented in Figure 3, and the resulting path decomposition in revised model is summarized as shown in Table 4.

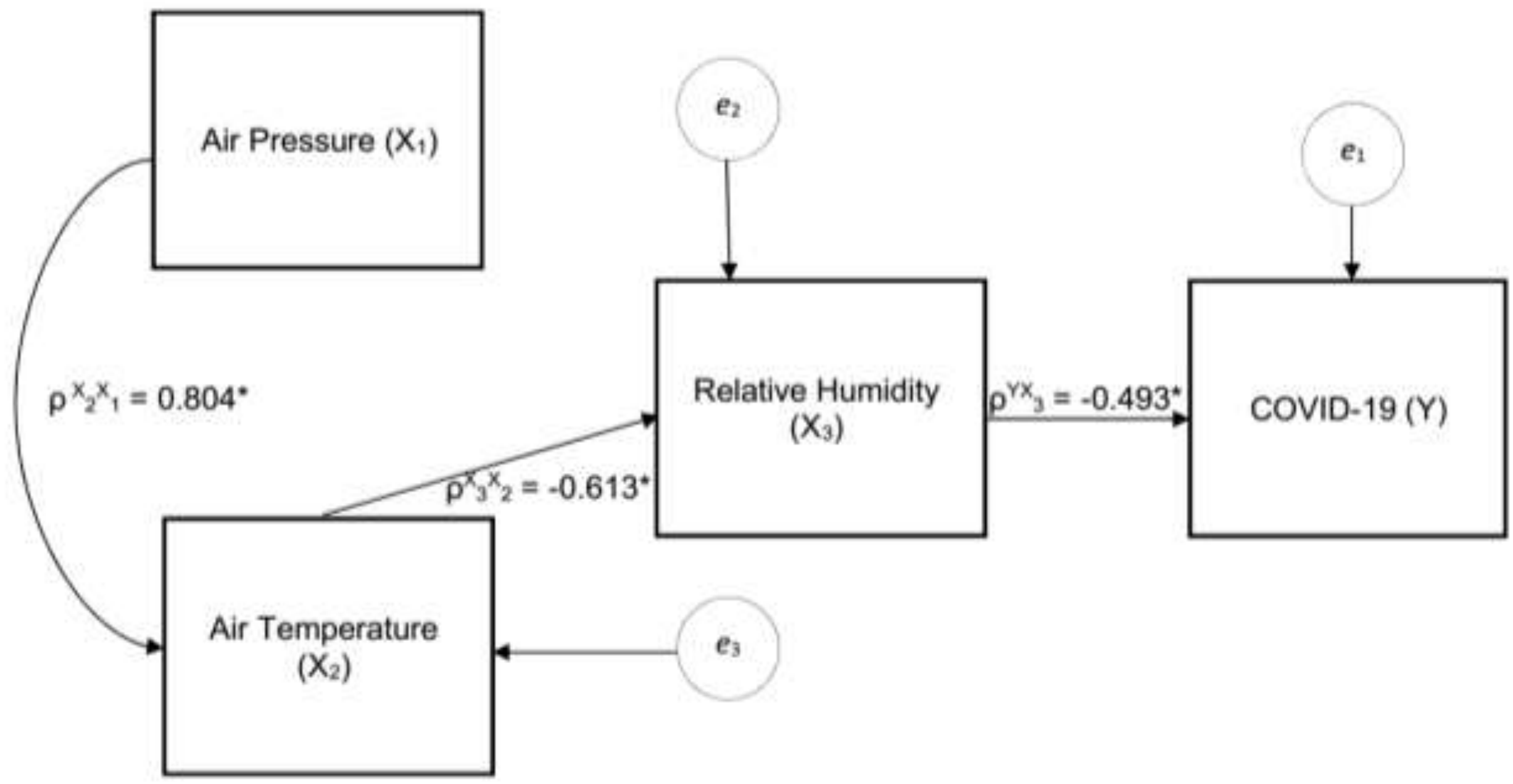

**Figure 3:-** Revised Causal Factor Models Affecting the COVID-19.

**Table 4:-** Calculation ofRevised Reproduced Correlation for the COVID-19 Model.

| Reproduced Correlation | Path Decomposition |
|---|---|
| $\hat{r}_{12}$ | $= \rho^{X_2X_1}$ |
| | = **0.804** |
| | (D) |
| $\hat{r}_{13}$ | $= (\rho^{X_2X_1})(\rho^{X_3X_2})$ |
| | = ( 0.804)(-0.613)= **-0.493** |
| | (I) |
| $\hat{r}_{1y}$ | $= (\rho^{X_2X_1})(\rho^{X_3X_2})(\rho^{YX_3})$ |
| | = (0.804)(-0.613)(-0.493)= **0.243** |
| | (I) |
| $\hat{r}_{23}$ | $= \rho^{X_3X_2}$ |
| | = **-0.613** |
| | (D) |
| $\hat{r}_{2y}$ | $= (\rho^{X_3X_2})(\rho^{YX_3})$ |
| | = (-0.613)(-0.493)= **0.302** |
| | (I) |
| $\hat{r}_{3y}$ | $= \rho^{YX_3}$ |
| | = **-0.493** |
| | (D) |

Once a model has been revised, the fit should again be reassessed in order to generate the best model. It can be gleaned from Table 5 that all of the reproduced correlations have no difference greater than 0.05 from observed correlation. Hence, those reproduced correlations indicate that the model is consistent with the empirical data.

**Table 5:-** Observed and Revised Reproduced Correlation for the COVID-19 Model.

| | $X_1$ | $X_2$ | $X_3$ | Y |
|---|---|---|---|---|
| Observed Correlation | | | | |
| $X_1$ | 1 | | | |

| $X_2$ | 0.804 | 1 | | |
|---|---|---|---|---|
| $X_3$ | -0.469 | -0.613 | 1 | |
| Y | 0.225 | 0.276 | -0.493 | 1 |
| Reproduced Correlation | | | | |
| $X_1$ | 1 | | | |
| $X_2$ | 0.804 | 1 | | |
| $X_3$ | -0.493 | -0.613 | 1 | |
| Y | 0.243 | 0.302 | -0.493 | 1 |

Based on the calculation of causal effects of the model presented in Table 6, only relative humidity was significantly related to the number of confirmed COVID-19 cases, and it can be observed that the coefficient path of relative humidity on inverse effect were good predictors of describingthe numbers of confirmed COVID-19 cases (p-values < 0.05). The table further showed that all factors explain for 24.5% of the variance on the number of confirmed COVID-19 cases; hence, about 75.5% could be attributed to the other factors not included in the study.

It was also observed that relative humidity was significantly influenced by air temperature which yielded an $R^2$ = 37.8%, and also air temperature was significantly influenced by air pressure yielded an $R^2$ = 64.7%.

**Table 6:-** Summary of Causal Effects for the number of COVID-19 confirmed cases.

| Outcome | Determinant | Causal Effects | | |
|---|---|---|---|---|
| | | Direct | Indirect | Total |
| Air Temperature ($R^2$ = **0.647**) | Air Pressure* | 0.804 | --- | 0.804 |
| Relative Humidity ($R^2$ = **0.378**) | Air Pressure | --- | -0.493 | -0.493 |
| | Air Temperature* | -0.613 | --- | -0.613 |
| COVID-19 ($R^2$ = **0.245**) | Air Pressure | --- | 0.243 | 0.243 |
| | Air Temperature | --- | 0.302 | 0.302 |
| | Relative Humidity* | -0.493 | --- | 0.493 |

## Conclusion and Policy Recommendation:-

The present study provided an empirical test of a causal model concerning relationships among air pressure, air temperature, relative humidity, and the number of confirmed COVID-19 cases. The results suggests that, as predicted, relative humidity has an inverse effect on number of confirmed COVID-19 cases, which indicate that the change in relative humidity affects the number of confirmed COVID-19 cases. Meanwhile, the change of air temperature affects the amount of relative humidity, and the variability in air pressure affects the amount of air temperature. In conclusion, the number of confirmed cases of COVID-19 may reduce as the amount of relative humidity increase, and relative humidity will increase, as the amount of air temperature decrease, and the amount of air temperature will decrease, as the amount of air pressure decrease.So, this path suggests, in order to forecast the number of confirmed cases in COVID-19, one might consider the causal ordering of the exogenous variables in order to describe clearly the number of confirmed COVID-19 cases. Hence, it is recommended that the government may increase the public awareness on the effects of causal ordering of weather condition, as it is one of the factor in order to control and forecast the number of confirmed COVID-19.

## References:-